\documentclass[11pt,oneside,a4paper]{article}

\usepackage{amssymb}
\usepackage{cite}
\usepackage{authblk}
\usepackage{graphicx}
\usepackage{caption}
\usepackage{wrapfig}
\usepackage{geometry}
\usepackage{amsmath}

\title{Bacterial chemotaxis: information processing, thermodynamics, and behavior}
\date{}

\author[1,2]{Gabriele Micali}
\author[1,2]{Robert G. Endres}
\affil[1]{Departement of Life Sciences, Imperial College, London, United Kingdom}
\affil[2]{Centre for Integrative Systems Biology and Bioinformatics, Imperial College, London, United Kingdom}

\begin{document}

\maketitle

\begin{abstract}

\textit{Escherichia coli} has long been used as a model organism due to the extensive experimental characterization of its pathways and molecular components. Take chemotaxis as an example, which allows bacteria to sense and swim in response to chemicals, such as nutrients and toxins. Many of the pathway's remarkable sensing and signaling properties are now concisely summarized in terms of design (or engineering) principles. More recently, new approaches from information theory and stochastic thermodynamics have begun to address how pathways process environmental stimuli and what the limiting factors are. However, to fully capitalize on these theoretical advances, a closer connection with single-cell experiments will be required.
\end{abstract}

\section*{Introduction}

All living organisms from animals to unicellular bacteria live under constant evolutionary pressure. To stay ahead in the game of evolution, organisms need to process noisy information, allowing them to make survival decisions quickly. However, to process information and move organisms also require energy. Thus the final behavior of any organism has to be an outcome which produces strong advantages under likely occurring environments. Chemotaxis of \textit{Escherichia coli} is particularly well understood in terms of its molecular components, allowing this bacterium to migrate towards food and away from toxins \cite{Parkinson15Rev,Tu13Rev,Sourjik12Rev,Armitage15Rev,Lai14Rev}. Indeed, an ever increasing amount of studies has highlighted several design principles, i.e. engineering blue prints, ensuring exquisite sensitivity, efficiency, robustness, and wide dynamic range at all levels of the pathway. 

This review focuses on recent findings in \textit{E. coli} chemotaxis, in particular on how molecular mechanisms give rise to information processing, its associated thermodynamic cost, and the resulting swimming behavior. What new design principles will be discovered next?

\section*{Classical view of \textit{Escherichia coli} chemotaxis}

\begin{figure}
  \includegraphics{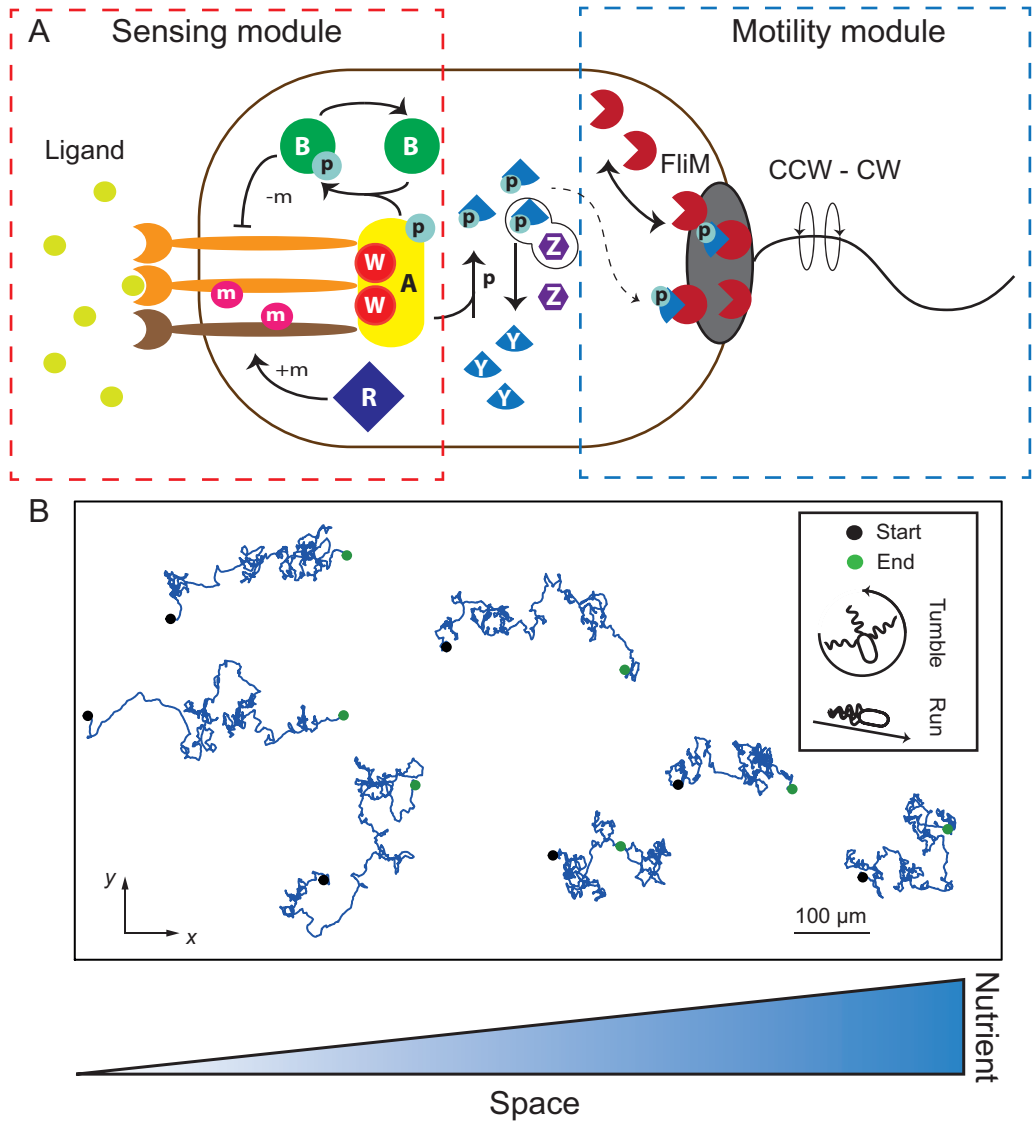}
  \caption{(A) Overview of chemotaxis. The sensing module (red box) includes different types of chemoreceptors grouped into clusters, kinase CheA (A), and adapter protein CheW (W). CheR (R) and CheB (B) regulate the methylation level of the receptors and hence memory, providing adaptation to persistent stimulation. The underlying mechanism of precise adaptation is integral feedback control as only inactive receptors are methylated by CheR and only active receptors are demethylated by CheB-p \cite{BarLei97}. The motility module (blue box) contains flagellar motors, which either rotate clockwise (CW) or counterclockwise (CCW). The two modules are linked by CheY (Y), which is phosphorylated by receptor-activated CheA (i.e. CheA-p) and dephosphorylated by phosphatase CheZ (Z). Once phosphorylated by ChaA-p, CheY-p diffuses in the cytoplasm to bind the FliM molecules of the motors, promoting CW rotation. (B) Simulations of swimming bacteria in a linear gradient of nutrient using RapidCell \cite{VlaLovSou08}, showing biased random walk.} \label{Fig1}
\end{figure}

\textit{Escherichia coli} is a Gram-negative bacterium inhabiting soil, as well as the animal and human gastrointestinal tracts. Inside the host, it contributes to the digestion of food and enhances resistance against pathogens \cite{NatInNat05}. This bacterium has a relatively simple chemotactic pathway (Fig. 1A). External stimuli are processed at the receptor level, where receptors sense and memorize chemical concentrations from the past by their adapted methylation level (Fig. \ref{Fig1}A, see red box for `sensing module'). The receptor-signaling activity can be monitored experimentally by tagging the CheY and CheZ proteins with a fluorescence-resonance-energy-transfer (FRET) reporter pair to follow their phosphorylation-dependent interaction. To swim cells are equipped with 5-8 flagellar rotary motors (Fig. 1A, blue box for `motility module'), each of which rotates either clockwise (CW) or counterclockwise (CCW). Taking together these flagella determine cell movement, given either by a `run' or a random reorientation in a `tumble' \cite{Darnton07JoB,Turner00JoB,Saragosti11}. The phosphorylated protein CheY-p links sensing and motility (Fig. 1A). In absence of any chemical gradient \textit{E. coli} performs random walk, and in a gradient, it biases its movement by having longer runs in favorable directions (Fig. 1B). 

Despite its simplicity, the chemotaxis pathway has astonishing properties. First, it is highly sensitive to detect small changes in chemical concentration. Indeed it works close to the physical limits of sensing \cite{BergPur77,Endres09PRL}, with a sensitivity equivalent to detecting as few as three molecules in the volume of the cell \cite{Mao03PNAS}. Second, \textit{E. coli}'s sensory system adapts to persistent stimulation \cite{BergTedesco75} and hence measures relative changes in ligand concentration rather than absolute concentrations (Weber's law) \cite{MesOrdAdl73}. Precise adaptation and signal amplifications by cooperative receptors allow \textit{E. coli} to perform chemotaxis over a wide range of concentrations \cite{DukeBray99,SouBerg02a,KeyEndSko06,EndWin06}. Furthermore, precise adaptation has been explained by a robust mechanism called integral feedback control \cite{BarLei97}, but recent experiments on swimming bacteria also demonstrate the limits of precise adaptation \cite{MassonCelaniVerg2012,Chemla12PNAS,NeumannNedVic14}. The adaptation mechanism provides bacteria with many additional advantages, some of which are known from neuroscience. When adaptation is precise the temporal profile of the response is largely independent of the strength of the stimulus (fold-change detection) \cite{ShovGoeAlon10,Sourjik15JoB}, and in the sensitive regime receptors perform logarithmic sensing (Weber-Fechner law) \cite{KalTuWu09}. Finally, the chemotaxis pathway is not only used to sense chemicals but also changes in temperature and pH \cite{OleksiukWinSou11,Sourjik12pH}. 

Despite these numerous findings, new directions of research have recently opened up in bacterial chemotaxis. Considering that the whole pathway is under evolutionary pressure, this renewed interest focuses on questions of optimality and trade offs in sensing and signaling. In the next sections we highlight recent findings about how the pathway processes information from the environment and the thermodynamic cost of performing chemotaxis.

\begin{figure}
  \includegraphics{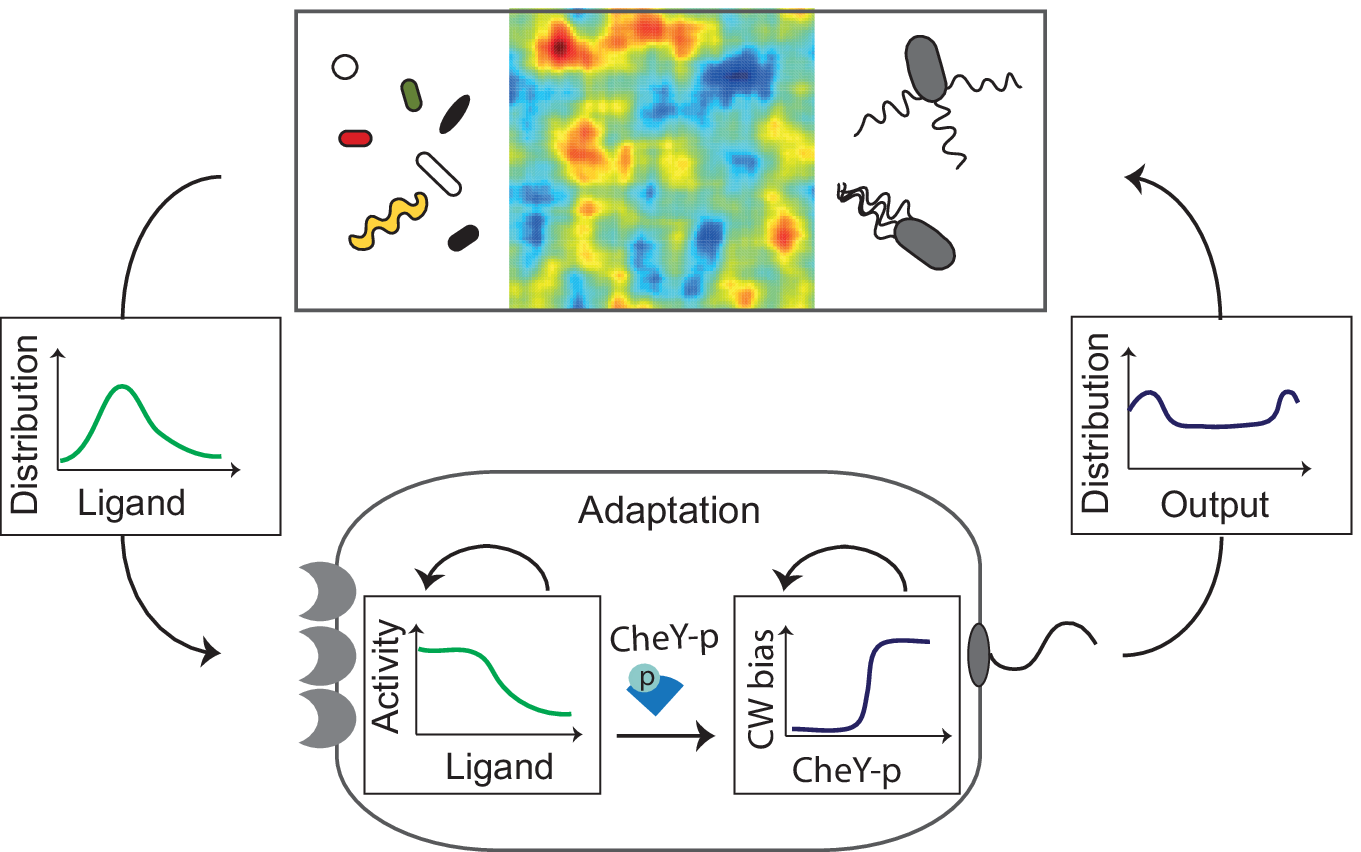}
  \caption{Bacterial chemotaxis under evolutionary pressure. Complex microenvironments (top) shape the ligand input distribution \textit{E. coli} experiences (left). Different bacterial species are indicated by different shapes and colors (top left), inhomogeneous chemical environments by colored patches (top centre), and bacterial motility by flagellated rods (top right).  This information is first processed by the receptors and finally by the motors (bottom). The final outcome (right) determinates the swimming behavior of the cell, and feeds back into the chemical environment and hence the inputs. Evolution has selected input-output relationships, which enhance the chance of the species' survival.} \label{Fig2}
\end{figure}

\section*{Information processing}

During their lives, cells experience many different inputs from the environment, to which they have to respond reliably with an output that enhances their chances of survival (Fig. \ref{Fig2}). Hence, cell-internal input-output relationships are shaped by evolution to transmit external information from the receptors to downstream proteins, which finally determine cell behavior. This information is inevitably corrupted by cell-external (extrinsic) and internal (intrinsic) sources of noise, and thus the best quantity for measuring information processing is Shannon's mutual information. Indeed, mutual information is a measure of statistical dependence between two random variables and describes the reduction of uncertainty of one after a noisy measurement of the other (see Box 1). But, have chemotactic cells evolved to maximize information transmission? 

The debate has just started. The assumption of maximal information transmission allows the reconstruction of distributions of optimal input and output distributions when dose-response curves and noise are measured at the receptor (or motor) level \cite{Micali14}. However, the complexity of the bacteria's microenvironments makes these predictions hard to test as bacteria are normally investigated under laboratory condition, not in their natural environment. More importantly, maximal mutual information at the sensing module maximizes the drift velocity of chemotactic cells in a linear gradient, linking mutual information to the final output of chemotaxis \cite{Micali14}. However, the ultra-steep dose-response curve of an individual motor \cite{CluSurLei00,YuanBerg13} might limit the transmitted information down to one bit of information (either CW or CCW rotation). To make things worse, the CheY-p concentration, which maximizes the drift, is in the saturated non sensitive tail of the motor dose-response curve \cite{Dufour14}. This implies non-optimal information transmission. However, \textit{E. coli} uses multiple flagellar rotary motors, which may increase information transmission at the very end of the pathway, e.g. by averaging out single-motor switching noise. Furthermore, the swimming trajectory should be regarded as the final output, constituting the spatial manifestation of all the information processed by the cell over some time. 

Despite these advances in understanding, the dependence of signaling on memory and hence the cell's individual history make the calculation of the mutual information between trajectories of inputs and outputs difficult \cite{TosttenWolde09}. Another trajectory-relevant technique is the k-space spatial information, shown to increase with gradient steepness \cite{Hoh12}. However, this spatial information is an image-analysis tool and not a quantification of information processing. Information processing also has its thermodynamic cost, which is conceptually difficult to understand as it relates to non-equilibrium physics.

\newpage

\subsection*{Box 1. Mutual information as a universal language of sensing and signaling.}

\renewcommand{\figurename}{ }
\begin{wrapfigure}{r}{0.5\textwidth}
  \includegraphics{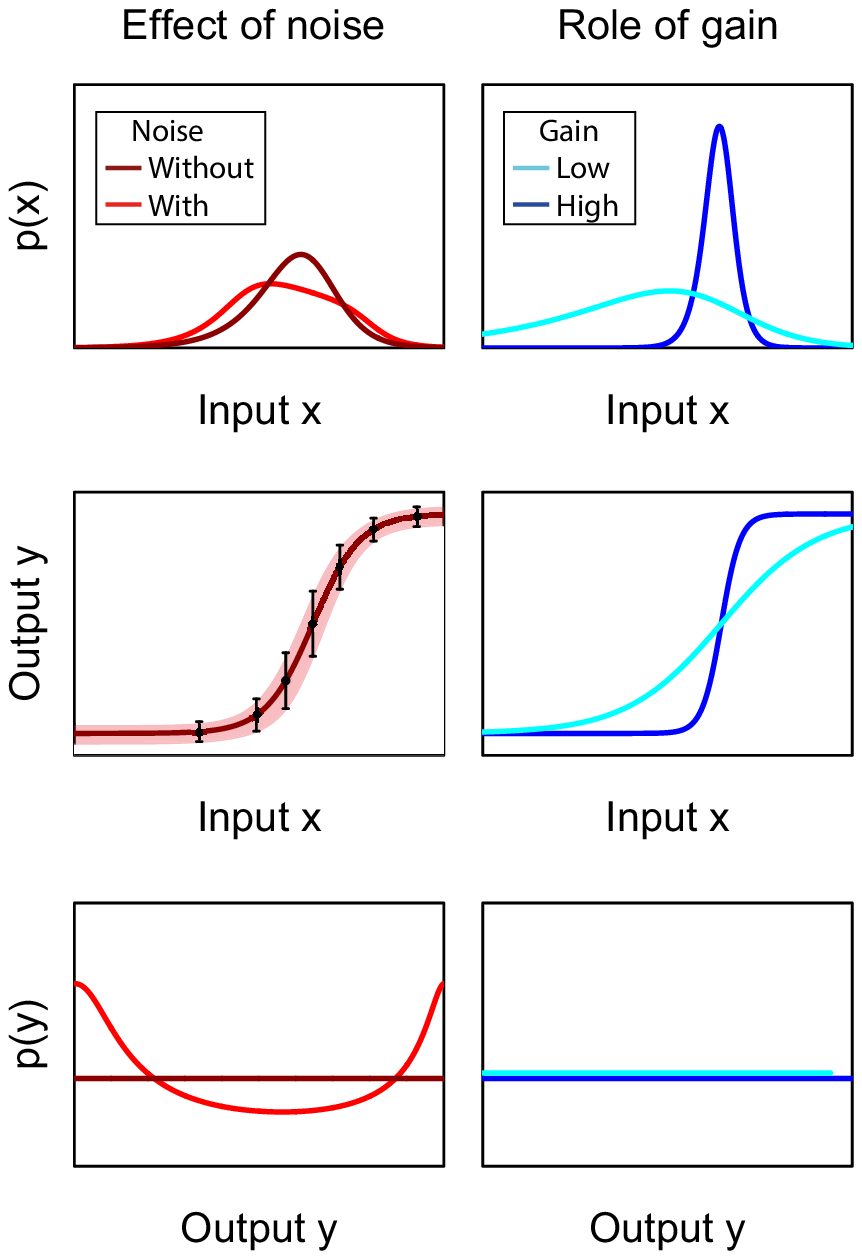}
\end{wrapfigure}

Information theory was first developed by Shannon in 1948 to quantify communication through a noisy channel \cite{Shannon}. Shannon defined a channel as a transmitting device: a message from a transmitter (input) is first encoded, then sent, and finally decoded by the receiver (output). In such a channel all steps may suffer from noise. The key quantity of Shannon?s theory is entropy (S), which measures the uncertainty of the outcomes of a random variable. Mutual information ($\mathcal{I}$) is defined as the reduction of uncertainty of the input knowing the output, and it is measured in bits. For continuous variables in a single processing step $(X\rightarrow Y)$, $\mathcal{I}$ is given by 
\begin{align*}     
 \mathcal{I}[X,Y] &=  S(X) - S(X|Y) \\
 &= \int_{X,Y} d x \, dy \, p(x,y) \log_2 \frac{p(x,y)}{p(x) p(y)} ,
 \end{align*}
where $S(X)$ is the Shannon entropy, i.e. the total uncertainty, of the input $X$; $S(X|Y)$ is the conditional entropy measuring the reduction of uncertainty after measuring $Y$; $p(x)$, $p(y)$ and $p(x,y)=p(y|x)p(x)$ are the probability distributions of input, output and joint probability, respectively. The conditional probability $p(y|x)$ is also called an input-output relationship.

Mutual information is indeed a measure of statistical dependence between two random variables. It accounts not only for linear dependencies (which for instance can be measured by Pearson correlation) but also for nonlinear dependencies \cite{Stumpf14Rev, Kraskov_04PRE}. Mutual information is independent of the units of measurement and details of the channel. It is symmetric and non-negative, i.e. is zero if and only if $x$ and $y$ are independent. Furthermore, information can only be lost by adding a processing step ($X\rightarrow Y\rightarrow Z$), never gained. However, measuring multiple inputs or outputs can in principle increase the mutual information. 

Due to the above mentioned properties mutual information seems `a' candidate to describe information processing between noisy environmental stimuli (input) and internal noisy protein levels (output) in biology. However, the most meaningful property of mutual information, which makes it `the' candidate, is its relation to decision theory and statistical inference: For a given input distribution, one input-output relationship allows better inference than another if and only if it has higher mutual information. (See \cite{SwainRevInfo14} for a review of information theory in biology).

For this reason, mutual information has recently been applied to experimentally measured dose-response curves of the receptor-signaling activity \cite{TkaCal08a}. When maximized given experimental dose-response curves and noise, it predicts optimal input, p(x), and output, $p(y)$, distributions. Input noise can produce a bimodal output distribution, e.g. in motor bias (left column in figure). A strongly bimodal distribution indicates that only one bit is transmitted in the pathway. A steep dose-response curve leads to high sensitivity and amplification of signals, but inputs may saturate the response, reducing the mutual information. Low gain might reduce the output range and thus also reduce the mutual information (right column in figure). Assuming that evolution favors high information transmission, selected dose-response curves might result from a trade off between high and low gain.

\section*{Thermodynamics of chemotaxis}

Enhancing information gain about the environment is generally evolutionarily favorable for cells (if not why having a sensory system?). However, nature needs to account for the thermodynamic cost of operating the biochemical signaling pathway. As found in neural coding \cite{Lau98,Lau81,Lau01}, evolution might search for an energy efficient way of transmitting information. Indeed, from the theoretical point of view, mutual information and energy dissipation of a system are fundamentally linked (see Box 2). Recent experiments at the nanometre scale demonstrate that computational devices must dissipate energy when performing the logically irreversible step of erasing information \cite{Berut12Nat}. Hence, the thermodynamic cost of chemotaxis needs to be understood. Does energy dissipation set a limit on the information that a cell can gain about its environment?

The debate touches on two main aspects of chemotaxis: the fundamental limit of sensing and the precision of adaptation. The fundamental limit of sensing, e.g. for inferring ligand concentration from inherently noisy measurements, is a theoretical upper bound on the accuracy a cell can reach \cite{BergPur77,Endres09PRL}. For instance, in seminal work Berg and Purcell used an equilibrium receptor but time averaging by an unknown ad-hoc mechanism \cite{BergPur77}. Hence, whether a cell can reach this limit is a different question since cells need to store the information in downstream proteins and implement time averaging to reduce noise, processes which generally require consumption of fuel molecules such as ATP and SAM. Indeed, theory shows that energetic cost limits the accuracy of sensing \cite{MehtaSchwab12,tenWolde14PNAS}. More specifically, there are three independent constraints to sensing, which cannot be traded off among each other: receptors and their averaging time, the number of downstream proteins, and fuel consumption \cite{tenWolde14PNAS}. The strategy of avoiding unfavorable bottle necks in the pathway, named optimal resource allocation, seems to be adopted by \textit{E. coli} \cite{tenWolde14PNAS}. However, these studies do not account for fluctuating environments, where the environment itself may provide work, enhancing information gain without internal fuel consumption \cite{BaratoSeifert14}. In \textit{E. coli} the cost of responding to a simple step-change in external ligand concentration is only about $10\%$ of the energy consumed for maintaining the dissipative steady state \cite{SartoriChiu14}. 

Why does nature adopt pathway architectures which lead to energy dissipation when there are potential pathways which do not consume fuel \cite{SartoriChiu14, DePalo13}? Indeed, there is a trade-off between energy consumption and fast, accurate adaptation \cite{DePalo13,LanTu12Nat}. Energy consumption can also reduce the correlations between extrinsic and intrinsic noise, ultimately resulting in a higher accuracy of sensing by time averaging \cite{Govern14PRL}. Finally, negative feedback during adaptation does not only reduce the adaptation error but also variability in adaptation \cite{Sartori15PRL}.

\subsection*{Box 2. Thermodynamics and information.}

The link between thermodynamics and information has a long history, which goes back to the famous thought experiment of Maxwell's demon. This fictitious demon can potentially produce order without doing any work, suggesting a violation of the second law of thermodynamics. However, to do so the demon has to measure and hence to acquire information about the environment. Performing a measurement moves the system out of equilibrium, and one needs to account for both the change in internal energy of the demon ($\Delta F^\text{H}$) and the change due to the measurement ($\Delta F^\text{meas}$, a change in information related to Shannon entropy and mutual information). It is then possible to fulfill the second law of thermodynamics and link energy dissipation ($W^\text{diss}$) with mutual information ($\mathcal{I}$):
\begin{align}
W^\text{diss} =T  \Delta S^\text{tot} = W - \Delta F^\text{H} - \Delta F^\text{meas} = W- \Delta F^\text{H} + k_B T \mathcal{I}  \ge 0 .
\end{align}
Moreover, information needs to be stored in `memories'. While it is possible to acquire information without energetic cost, the erasure of information (the demon is a finite being) must dissipate energy (Landauer's principle). (See \cite{ParHorowSagawa15} for a review on thermodynamics of information.)

The generality of these arguments suggests that a measurement, and hence acquisition of information, is linked to energy dissipation independent of the underlying mechanism. In \textit{E. coli}, for instance, information about the extracellular concentration of chemicals is `stored' in the adapted methylation level of receptors. Once the external concentration changes, the receptor activity responds rapidly, enhancing its correlations with the environment. At the same time, the methylation level still contains the old information, i.e. reflects the previous ligand concentration. Subsequently, during adaptation, the methylation level adjust and thus acquires new information while the activity returns to the adapted value, which is independent of the environment. While \textit{E. coli} even consumes energy at the adapted steady state, the response to an environmental change is also an out of equilibrium process \cite{SartoriChiu14}.

\section*{Cell-to-cell variability and cell behavior}

The search for general design principles is critical for simplifying molecular complexity and hence for understanding biological systems. All previously discussed design principles describe single cells, e.g. about how they deal with signaling noise. However, noise also leads to cell-to-cell variability, affecting populations of cells. For instance, there is large variation in adaptation times probably due to low copy numbers of CheR and CheB \cite{LiHaz04, VlaLovSou08}. In addition, different adapted CheY-p levels have been observed in cells of the same population \cite{AlonSurLeib99}. Hence, the tumble bias, which is determined by the CheY-p level, also differs from cell to cell. Furthermore, cells vary in total number and spatial distribution of receptors \cite{ThiemSourjik08} and motors \cite{Cohen08}. All this variability might be helpful under unpredictable changes of the environment, where having different phenotypes increases the chances of a species' survival \cite{FrankelEmonet14Elife}. By focusing on cell-to-cell variability previously hidden principles might emerge.

How do \textit{E. coli} cells tune the ultra-steep dose-response curves of the motor to their individual adapted CheY-p level? This question is of particular importance because these curves are even steeper than previously thought (Hill coefficient up to 20) \cite{YuanBerg13}. Recent, experiments show that even mutants with non-adapting receptors can partially adapt by exchanging FliM molecules between the cytoplasm and the motor (Fig. 1A) \cite{YuanBerg12Nat}. By changing the number of FliM in the motor cells adjust the sensitive regime of their motors to the cell-specific adapted CheY-p level \cite{YuanBerg12Nat}. However, this adaptation mechanism is not regulated by CheY-p directly. A change in binding affinity between FliM and the motor substrate due to changes in the rotational motor bias seems to guide motor adaptation \cite{LeleBerg12PNAS}. 

Finally, optical traps allow long-time measurements of individual cells in response to external changes in ligand concentration \cite{MinChemla09}. This technique helped to better understand the large variation in number of motors in \textit{E. coli} \cite{Chemla14Elife}. In fact, coupling between motors rotating in the CW direction, e.g. through their flagella, results in a tumble bias, which is robust against the number of motors, a design principle previously missed \cite{Darnton07JoB,Chemla14Elife,SneddonEmonet12,HuTu13}. 

\section*{Conclusions}

The results reported here have been possible due to new theoretical approaches, in particular in information theory and stochastic thermodynamics. Despite their long history, only recently have they been used in biology. However, most of the results obtained lead to predictions that necessarily need to be tested experimentally in the future. The focus will inevitably be on shifting from physics-type experiments to investigations of living cells. Another goal will be the characterization of trajectories of swimming bacteria, and hence of individual cell behavior. Commercially available high-resolution imaging is already a valuable tool to study variability among different cells. Moreover, optical tweezers allow for measurements on single cells, suitable for studying temporal variability in single cells over an extended period of time. Combining internal measurements of signaling in individual cells, e.g. by single-cell FRET, with history-dependent cell trajectories of behavior will open new doors even in the best-known system of cell biology.

\section*{Acknowledgments}

We thank Nikola Ojkic and Chiu Fan Lee for critical reading of the manuscript. This work was supported by European Research Council Starting-Grant N. 280492-PPHPI.

\bibliographystyle{plos2009}
\bibliography{all}

\end{document}